# Using ChatGPT-4 for the Identification of Common UX Factors within a Pool of Measurement Items from Established UX Questionnaires

Stefan Graser, Stephan Böhm

CAEBUS Center of Advanced E-Business Studies
RheinMain University of Applied Sciences
Wiesbaden, Germany
e-mail: {stefan.graser, stephan.boehm}@hs-rm.de

Martin Schrepp

SAP SE
Walldorf, Germany
e-mail: martin.schrepp@sap.com

*Abstract*—Measuring User Experience (UX) with standardized questionnaires is a widely used method. A questionnaire is based on different scales that represent UX factors and items. However, the questionnaires have no common ground concerning naming different factors and the items used to measure them. This study aims to identify general UX factors based on the formulation of the measurement items. Items from a set of 40 established UX questionnaires were analyzed by Generative AI (GenAI) to identify semantically similar items and to cluster similar topics. We used the LLM ChatGPT-4 for this analysis. Results show that ChatGPT-4 can classify items into meaningful topics and thus help to create a deeper understanding of the structure of the UX research field. In addition, we show that ChatGPT-4 can filter items related to a predefined UX concept out of a pool of UX items.

*Keywords–User Experience (UX); UX Measurement; UX Factors; Measurement Items; Generative AI (GenAI); Large Language Model (LLM); ChatGPT; Semantic Textual Similarity (STS).*

## I. INTRODUCTION

User Experience (UX) is a holistic concept in Human-Computer-Interaction (HCI) describing the perception towards the use and interaction of a product, service, or system [1]. A positive UX is essential for interacting with products and services. This user's perception must be considered to gather insights into improving the UX [2]. Therefore, various methods can be found for UX measurement. The most common way to measure the UX is through standardized questionnaires providing self-reported data by the user [3]. These questionnaires can be applied in a cost-efficient, simple, and fast way [3][4].

Over the last decades, different standardized questionnaires were developed, breaking down and measuring the construct of UX. Therefore, the questionnaires refer to a holistic view or focus on a specific dimension. To be more precise, a questionnaire is based on the different factors, items, and scales about the respective dimension [5][6]. However, there is no common ground within the factors and items among the standardized UX questionnaires. Differently named factors can measure the same, but factors with the same name can measure something different [7]. This leads to a blurring of the respective measurement focus among the questionnaires. Nevertheless, a clear distinction between the measurement items is necessary to measure the same and have a shared understanding of the construct of UX. There is a lack of sufficient exposition of what different developed scales semantically mean [7].

In this regard, this study focuses on the level of the different items describing the UX dimensions. We aim to identify semantically similar items by applying Generative AI. Therefore, we used ChatGPT-4 as a Large Language Model (LLM) to analyze and compare items concerning their Semantic Textual Similarity (STS). Based on this, similar items were clustered. As a result, we try to identify UX topics from these clusters. Against this background, we address the following research questions:

**RQ1:** *Is Generative AI able to identify useful similarity topics based on measurement items?*

**RQ2:** *Which topics based on semantically similar measurement items can be identified among the most established UX questionnaires?*

This article is structured as follows: Section 2 describes the theoretical foundation of this approach. Section 3 shows related work concerning the consolidation of UX factors and common ground in UX research. Section 4 illustrates the methodological approach by applying the LLM ChatGPT-4 as Generative AI. Results are shown in Section 5. A conclusion and outlook is given in Section 6.

## II. THEORETICAL FOUNDATION

### A. Concept of UX

As already described, UX is a multidimensional construct consisting of different dimensions and quality aspects. Usability, which is defined as "the extent to which a product can be used by specified users to achieve specified goals with effectiveness, efficiency, and satisfaction in a specified context of use" [1] is focused on completing tasks and achieving goals. UX, on the other hand, encompasses a broader spectrum of qualities related to a product's subjective impression. This includes, for example, aspects such as aesthetics or fun of use. Thus, usability can be declared a subset of UX [8].

Based on this, Hassenzahl [9] presents a distinction between pragmatic and hedonic properties. Pragmatic qualities are task-related, whereas hedonic qualities refer to non-task-related qualities [9]. However, this distinction is accompanied by problems. Firstly, a clear distinction is not always possible for a specific product. Secondly, pragmatic qualities relate to a common concept as they are task-related whereas hedonic qualities do not follow such a concept [6].

Schrepp et al. [6] followed a new approach conceptualizing UX as a defined set of quality aspects. A "UX quality aspect describes the subjective impression of users towards a semantically clearly described aspect of product usage or product design" [6]. This results in clearly described and distinct aspects that can be used to evaluate the subjective experience towards a product [6].

*B. Semantic and Empirical Similarity*

In this paper, we focus on investigating the semantic similarity of measurement items from UX questionnaires. Semantic similarity refers to the degree of likeness or resemblance between the item texts based on their meaning. Thus, semantic similarity expresses how closely related the underlying textual concepts are, rather than just the surface-level syntactic or structural similarity. Semantic similarity takes into account the context, relationships, and associations between words or phrases to determine their level of similarity [10]–[12]. Different statistics-based methods in Natural Language Processing (NLP) for Semantic Textual Similarity measurement can be found in the literature [10][13]–[20]. In general, the methods can be divided into the three categories Matrix Based Methods, Word Distance-Based Methods, and Sentence Embedding Based Methods [21].

Large Language Models, like GPT, use word embeddings (dense vector representations of words derived with the help of deep learning mechanisms applied to vast volumes of existing texts) to calculate semantic similarity. Thus, they are obviously helpful tools for analyzing the semantic similarity of UX items.

However, in interpreting the results of such an analysis of semantic item similarity, we must distinguish the semantic similarity of items from their empirical similarity [22][23], i.e., their empirical correlation, to understand the benefits and limitations of such an approach. We may observe items that have a small semantic similarity as estimated by an LLM but show in empirical studies quite substantial correlations.

A well-investigated example is the observation that beautiful products are perceived as usable [24][25]. Thus, visual aesthetics influence the perception of classical UX aspects like *Efficiency*, *Learnability*, or *Controllability*, and items measuring these semantically quite different aspects correlate. A similar effect exists also in the opposite direction, i.e., the perception of *Usability* influences the perception of beauty [26][27].

There are several explanations (which in fact may all contribute to the effect) for such first-sight strange empirical dependencies, for example, the general impression model [28], evaluative consistency [29], or mediator effects [30]. Another explanation is that aesthetics and usability share, in fact, some common aspects. Balance, symmetry, and order [31] or alignment [32] influence the aesthetic impression. But a UI that looks clean, ordered, and properly aligned is also easy to scan and thus, users can find elements faster and orient more easily on such an interface. Hence, it will also benefit *Efficiency* or *Learnability* [23].

Given these arguments, we can expect that items with a high semantic similarity will also show empirically high correlations (they ask for highly similar UX aspects thus, participants of a survey should give highly similar answers). However, there may be items with quite low semantic similarities but quite high empirical correlations due to the effects described above. Thus, we should not expect that we can reconstruct typical scales of established questionnaires by a purely semantical analysis of the items. Such scales are usually developed by an empirical process of item reduction, mostly by main component analysis and group items based on empirical correlations from larger studies.

*C. UX Questionnaires*

Quantitative UX evaluation is usually based on questionnaires as subjective assessments of user's perceptions. Various standardized UX questionnaires can be found in scientific literature. For example, Schrepp [7] describes 40 quite common UX questionnaires [7]. Every questionnaire is based on specific factors, items, and scales. Moreover, measurement focus can differ among the questionnaires. The selection of the specific questionnaire may differ depending on the application purpose or objective of the investigation.

Díaz-Oreiro et al. [33] investigated the User Experience Questionnaire UEQ [34] as the most widely used questionnaire for UX evaluation. This can be confirmed by further research [33]. The UEQ developed by Laugwitz et al. [34] is based on the UX framework by Hassenzahl [9][34]. The questionnaire consists of six factors divided into pragmatic and hedonic properties. Each factor contains four items formulated as a semantic differential scale measured by a 7-point Likert scale. The factors with their descriptions are shown below:

- **Attractiveness**: Overall impression of the product. Do users like or dislike it?
- **Perspicuity**: Is it easy to get familiar with the product and to learn how to use it?
- **Efficiency**: Can users solve their tasks without unnecessary effort? Does it react fast?
- **Dependability**: Does the user feel in control of the interaction? Is it secure and predictable?
- **Stimulation**: Is it exciting and motivating to use the product? Is it fun to use?
- **Novelty**: Is the design of the product creative? Does it catch the interest of users?

The questionnaire aims to gather a holistic impression referring to the UX of interactive products. The UEQ is an example of a questionnaire with scales representing quite abstract UX concepts and can thus be applied to many different products. The items are semantic differentials, i.e., pairs of terms with opposite meanings that represent a semantic scale (for example, slow/fast). Further details can be found online [35].

Other established questionnaires follow a different measurement concept in that their items and scales refer to concrete interface elements. For example, the Purdue Usability Testing Questionnaire [36] contains items like "Is the cursor placement consistent?" or "Does it provide visually distinctive data fields?". This form of items is much more concrete but can only be applied to a certain type of product. In addition, there are several questionnaires that can be applied only for special application domains, for example, web pages, e-commerce, or games (for an overview of common questionnaires and item formulations, see [37]). This huge variety in the way items are formulated makes it also quite challenging to categorize them concerning their semantic meaning.

No questionnaire can cover all UX factors. As already described, each questionnaire refers to a specific focus. Therefore, it is a common way to combine or apply several questionnaires simultaneously to cover all relevant aspects. Due to different items and scales, it may be more difficult for participants to complete the evaluation. Therefore, Schrepp and Thomaschewski (2019) developed the UEQ+, a modular framework. The framework is based on described factors with their respective items covering the construct UX as broadly as possible. Researchers can choose from a set of 16 UX quality aspects according to the respective product to evaluate and create an individualized UX questionnaire [38]. Further information can be found online [39].

### III. RESEARCH OBJECTIVE AND RELATED WORK

Due to the high number of UX questionnaires developed in the last decades, many different factors and items can be found. This emphasizes the lack of common ground within quantitative UX evaluation. Concerning this research gap, only a little research was done to consolidate general UX factors and find a common understanding.

[40] aimed to consolidate a list of general UX factors. Therefore, existing questionnaires and literature were analyzed. All collected factors were then consolidated based on their definition. This resulted in a consolidated list of general UX factors [40]. The same approach was conducted by [5] and [6]. The latest list of consolidated UX factors is shown in the following table (see Table I):

Typically, UX factors are constructed with the help of empirical methods of item reduction, for example, main component analysis. Thus, items are grouped into factors based on their empirical correlations. This leads sometimes to scales that consist of items that represent, at least at first sight, semantically different concepts. Thus, it is sometimes difficult to clearly describe what the semantic behind a scale actually is. To get a deeper understanding of the concept of UX, it makes thus sense to analyze the purely semantic similarities of items and to investigate a structuring based on this concept.

Only two studies have yet applied methods to measure semantic textual similarity in the field of UX research concerning UX measurement items. Both studies applied NLP techniques at the level of the measurement items. In particular, the semantic textual similarity between the measurement items was analyzed. By doing this, the researchers tried to ensure a more accurate distinction. In particular, a Sentence Transformer Model and a Sentence Transformer-based Topic Modeling approach were conducted concerning the semantic structure of the textual items [41][42].

The first study by [41] applied the Sentence Transformer Model Augmented SBERT (AugSBERT) [20] to measure the sentence similarity using a cross- and bi-encoder Transformer architecture to encode the measurement items of established UX questionnaires into embedding in a vector space. Afterward, the cosine similarity values between the items were calculated and items were clustered based on a determined threshold. As a result, the similarity clusters containing semantically similar items were identified [41]. The second study extends this approach by applying the specific Topic Modeling technique BERTopic [43] based on the Sentence Transformer SBERT [19]. Therefore, the items were encoded into embeddings in a vector space by applying the SBERT approach. Moreover, the embeddings were clustered using a Topic Modeling technique [42]. Both studies show that innovative NLP techniques can produce plausible results. Nevertheless, there are still several weaknesses in the approaches to be recorded. For further insights, we refer to the respective articles [41][42].

TABLE I: CONSOLIDATED UX FACTORS BASED ON [6].

| (#) | Factor | Descriptive Question |
|---|---|---|
| (1) | Perspicuity | Is it easy to get familiar with the product and to learn how to use it? |
| (2) | Efficiency | Can users solve their tasks without unnecessary effort? Does the product react fast? |
| (3) | Dependability | Does the user feel in control of the interaction? Does the product react predictably and consistently to user commands? |
| (4) | Usefulness | Does using the product bring advantages to the user? Does using the product save time and effort? |
| (5) | Intuitive use | Can the product be used immediately without any training or help? |
| (6) | Adaptability | Can the product be adapted to personal preferences or personal working styles |
| (7) | Novelty | Is the design of the product creative? Does it catch the interest of users? |
| (8) | Stimulation | Is it exciting and motivating to use the product? Is it fun to use? |
| (9) | Clarity | Does the user interface of the product look ordered, tidy, and clear? |
| (10) | Quality of Content | Is the information provided by the product always actual and of good quality |
| (11) | Immersion | Does the user forget time and sink completely into the interaction with the product |
| (12) | Aesthetics | Does the product look beautiful and appealing? |
| (13) | Identity | Does the product help the user to socialize and to present themselves positively to other people? |
| (14) | Loyalty | Do people stick with the product even if there are alternative products for the same task |
| (15) | Trust | Do users think that their data is in safe hands and not misused to harm them? |
| (16) | Value | Does the product design look professional and of high quality? |

Due to the rapid development of Generative AI, various fields, e.g., NLP are revolutionized [44][45]. Therefore, Generative AI (GenAI) is able to improve processes and contribute valuable results. This study is another approach applying GenAI to find common ground in UX research. We used ChatGPT-4 as LLM [46] to clearly differentiate items semantically and consolidate general factors within established UX questionnaires. The detailed approach is explained in the following section IV.

### IV. METHODOLOGICAL APPROACH

This study applies GenAI for the analysis of UX measurement items. In particular, ChatGPT-4 was used to determine similarity topics based on semantically similar items. The approach is described in the following. ChatGPT-4 is a large multimodal model developed by OpenAI that is able to process data and produce text outputs. The model based on GPT-4 is capable of understanding and generating natural language text [46]. For detailed insights, we refer to OpenAI (https://openai.com/gpt-4).

As a first step in our approach, data was collected. A set of 40 established UX questionnaires [7] was analyzed. We excluded all questionnaires with (1) a semantic differential scale and (2) a divergent measurement concept, i.e., specifically formulated items focusing on a concrete evaluation objective (for further details, see section II-C). This resulted in a list of 19 questionnaires with 408 measurement items. The data collection process is illustrated in Figure 1.

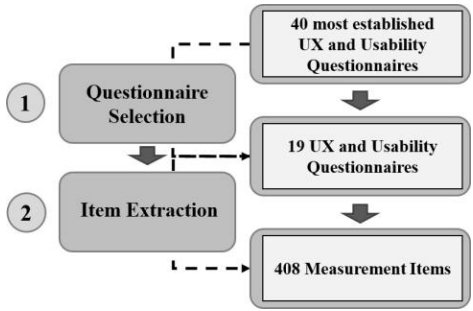

Figure 1: Data Collection.

Secondly, we introduced all items to ChatGPT-4. Thirdly, we formulated seven prompts for ChatGPT-4. The prompts described the task for the LLM. The different tasks given to ChatGPT are described in detail below. The prompts are shown in the following:

- **prompt1**: *"Can you extract the questions with a high similarity, i.e., answering about similar topics?"*
- **prompt2**: *"Can you break this down more detailed?"*
- **prompt3**: *"Can you try to break down each section into more subsections with its own category?"*
- **prompt4**: *"Can you improve your categorization?"*
- **prompt5**: *"In literature, I can find such a list with 16 UX factors.—inserted the defined quality aspects (see Table I)—. Can you compare this list with your categorization and contrast these lists?"*
- **prompt6**: *"I would like you to take your categorization you have done earlier and improve this into more generalized, holistic topics"*
- **prompt7**: *"Below there is a list of statements and questions related to the UX of a software system. Select all statements or questions from this list that describe how easy or difficult it is to learn and understand how to use the software system. List these statements or questions. Start with those statements and questions that describe this best.—inserted list of 408 items from UX questionnaires."*

In relation to *prompt1*, a simple classification was performed. Based on this, *prompt2* should be used for a first extension and development of specific topics. In the next step, the topics were further divided into subcategories *prompt3*. Further, with *prompt4* the task of a topic improvement was specified. For this, the LLM should try to optimize the topics and the respective subcategories classified so far and, thus, create a further advanced classification. Finally, existing UX quality aspects from the literature were introduced to ChatGPT and compared with the AI-generated topics in relation to their similarities and differences *prompt5*. Until now, we made an exploratory structuring. Moreover, we want ChatGPT to generate and improve the categorizations into more general topics providing a holistic perspective *prompt6*. Furthermore, we aimed to filter out suitable items that fit a category very well from the existing set of items. We set *prompt7* to detect appropriate items using the example of the UX quality aspect ***Learnability***. Such detecting and assignment is particularly useful for "ad-hoc surveys" that do not use a standardized questionnaire to measure UX, but just a bunch of self-made questions to find out something specific. This often requires spontaneous additional questions. Thus, before formulating new items, the search and detection of measurement items within an existing item pool using GenAI is quite practical. Results are shown in the following Section V.

## V. RESULTS

In this section, the results of the approach by applying ChatGPT-4 are shown. The sub-sections are aligned to the respective prompts that have been given to ChatGPT.

### A. Prompt1: Primary Classification

Referring to the first prompt, the LLM provided a classification by themes and similar topics. This results in six topics. Additionally, the most suitable items have been assigned to each topic. Due to paper restrictions, we have only provided the first three most representative items listed by ChatGPT for each category (see Appendix A1). The classification is shown in the following:

- (1) **Usability and Ease of Use**
- (2) **Design and Aesthetics**
- (3) **User Engagement and Experience**
- (4) **Trust and Reliability**
- (5) **Information Access and Clarity**
- (6) **Issues and Errors**

With regard to the results, common topics emerge. Therefore, functional as well as emotional topics were generated. While observing the items, the topics with their respective items can be considered plausible. Concerning the items, it must be pointed out that the item formulations are very specific, while the different categorizations are very broad in comparison. For example, Topic (1) is named **Usability and Ease of Use**, but the first three representative items refer specifically to Ease of Use. Thus, the respective topics are very broad.

The LLM can identify logical topics based on the semantic textual structure. Nevertheless, as a classification of 6 topics with a total of 408 items seems very superficial, we directly proceeded to the next step. Here we asked the LLM for a more specific classification.

### B. Prompt2: More Detailed Classification

We tried to derive a more detailed classification. The respective items are presented in the Appendix (see A2). As a result, ten topics were determined by the LLM.

- (1) **Ease of Use**

- (2) **Complexity and Usability Issues**
- (3) **Design and Appearance**
- (4) **Engagement and Immersion**
- (5) **Performance and Responsiveness**
- (6) **Reliability and Trust**
- (7) **Information Quality and Access**
- (8) **Errors and Bugs**
- (9) **Learning and Memorability**
- (10) **Effectiveness and Efficiency**

Considering the results, the second classification is more precious containing four more topics. Topic (1) in relation to *prompt1* was divided into two topics. Additionally, Performance and Responsiveness, Learning and Memorability, and Effectiveness and Efficiency were introduced. By comparing the results of the first two prompts, the functional, task-related topics were further broken down. Thus, the LLM can distinguish the topics even more precisely. It can be seen that the majority of the AI-generated topics relate to a rather pragmatic quality. Topic (1), (2), (5), (7), (8), (9), and (10) are task-related whereas (3) and (4) address the emotional perception of the user. Topic (6) – Reliability and Trust – contains both task-related and emotional items. Overall, the measurement items seem to be more functionally driven among the topics. Moreover, the item formulation within the different topics is quite broad. Some items can be applied to many scenarios, e.g., *"it meets my needs"*, while others are specified to an application, e.g., *"I feel comfortable purchasing from the website"*. An even more detailed categorization into subcategories therefore seems reasonable.

*C. Prompt3: Extended Classification*

We tried to provide a more detailed classification within each topic and asked for a specific breakdown into subsections. As a result, we obtained 22 further subtopics:

- **Ease of Use**
  System Usability—Website Usability—Application Usability
- **Complexity and Usability Issues**
  System Complexity—Frustration and Difficulty—System Limitations
- **Design and Appearance**
  Visual Attraction—Layout and Structure—Design Consistency
- **Engagement and Immersion**
  Time Perception and Involvement—Depth of Experience
- **Performance and Responsiveness**
  Speed of Response
- **Reliability and Trust**
  Website Trustworthiness—System Reliability
- **Information Quality and Access**
  Quality of Information—Accessibility of Information
- **Errors and Bugs**
  Technical Issues—Error Messages
- **Learning and Memorability**
  Learning Curve—Recall and Retention
- **Effectiveness and Efficiency**
  Functional Efficiency—Expected Functionality

The division into main topics and respective sub-topics confirms that items have the same characteristics on a higher level, but can be further subdivided on a more specific level. This may be due to the different characteristics and focus of the questionnaires and their items. Up to this point, we have determined what categorization levels ChatGPT should take. The next step is to extend ChatGPT to make improvements within its own categorization.

*D. Prompt4: Classification Improvement*

We want ChatGPT to improve the classification without any further specifications. As a result, the LLM identified six main topics with 16 subtopics. For improvement, the number of main topics was reduced which makes it appear that the main topics are again rather broad. This results as well in a broad spectrum of sub-topics. Within the sub-topics, ChatGPT changed the categorizations. For instance, hedonic categories, e.g. **Aesthetics and Design**, are grouped with pragmatic categories, e.g. **Navigation and Usability**. In contrast, the main topic **System Usability and Performance** contains the three sub-topics **Ease of Use, Efficiency and Speed, and Functionality and Flexibility**. Compared to the definition by the DIN ISO [1], the concept of usability is mostly well captured. Concerning the properties, more topics are functional than emotional.

- **System Usability and Performance**
  Ease of Use—Efficiency and Speed—Functionality and Flexibility
- **User Engagement and Experience**
  Engagement Level—Aesthetics and Design—Confusion and Difficulty
- **Information and Content**
  Clarity and Understandability—Relevance and Utility—Consistency and Integration
- **Website-specific Feedback**
  Navigation and Usability—Trust and Security—Aesthetics and Design
- **Learning and Adaptability**
  Learning Curve—Adaptability
- **Overall Satisfaction and Recommendation**
  Satisfaction—Recommendation

In consideration of the results, the categorization improvement emphasizes the two-level structure of the main and sub-topics. However, some main topics are rather broad containing sub-topics with pragmatic as well as hedonic properties.

*E. Prompt5: Comparison Towards Existing Consolidation*

In the following step, we consulted existing UX concepts (see Table I) developed by [6] and compared them to the AI-generated categories. We attempted to draw a comparison between an existing consolidation and the results of the LLM. We defined the prompt as follows: *"In literature, I can find such a list with 16 UX factors.—inserted the defined quality aspects (See Table I) [6]—. Can you compare this list with your categorization and contrast these lists?"*. The comparison is illustrated in Table II:

TABLE II: COMPARISON OF EXISTING UX QUALITY ASPECTS [6] AND AI-GENERATED TOPICS.

| (#) | UX Quality Aspects | AI-generated Sub-Topics |
|---|---|---|
| (1) | Perspicuity | Ease of Use—Learning Curve |
| (2) | Efficiency | Efficiency and Speed |
| (3) | Dependability | Consistency and Integration |
| (4) | Usefulness | Functionality and Flexibility—Relevance and Utility |
| (5) | Intuitive use | Ease of Use |
| (6) | Adaptability | Adaptability |
| (7) | Novelty | - |
| (8) | Stimulation | Engagement Level |
| (9) | Clarity | Clarity and Understandability |
| (10) | Quality of Content | Relevance and Utility |
| (11) | Immersion | Engagement Level |
| (12) | Aesthetics | Aesthetics and Design—Aesthetics and Design |
| (13) | Identity | - |
| (14) | Loyalty | Loyalty |
| (15) | Trust | Trust and Security |
| (16) | Value | Perceived value |

In relation to this comparison, ChatGPT shows some fundamental differences. Firstly, an allocation of the AI-generated topics to all quality aspects is not possible. The factors of *Novelty* and *Identity* stated in the literature [5][6][40] are not covered in the categorization made by ChatGPT. Moreover, there is some overlap between the items and factors as some AI-generated factors can be allocated to more than one quality aspect. Furthermore, the results of the literature (see Table I, [6]) are more generalized. For example, the sub-topic *Trust and Security* is contained in the main topic *Website-specific Feedback*. Hence, *Trust and Security* refers specifically to Websites. In contrast, the quality aspect of *Trust* defined by Schrepp et al. [6] is a main topic of its own described more generally. Thus, existing quality aspects introduce a more holistic view covering both functional and emotional aspects of UX whereas the categorization of the LLM has a stronger focus on the functional side and is more specific. If the categories are too specific, there may be problems with general applicability. Therefore, the objective remains to formulate and present (1) more generally and (2) more emotionally focused categories to provide a universal and holistic perspective towards UX.

*F. Prompt6: Construction of Generalized Categories*

Against this, we added a further prompt *"I would like you to take your categorization you have done earlier and improve this into more generalized, holistic topics"* to create more generalized topics. In this regard, it is also important to see which items represent the generated topics according to the GenAI as the consolidation and categorization are originally based on the measurement items. We output the top five items representing the respective topic best. As a result, ChatGPT generates a comprehensive overview with generalized UX factors and their definitions. The classification shows a two-dimensional separation into the main topic and sub-topics. Both functional, task-related as well and emotional aspects are contained. This enables a comprehensive and generalized view of the construct of UX made by ChatGPT. The topics and items are shown in the appendix (see A3).

Considering the results, ChatGPT performs very well in consolidating and developing topics concerning a holistic view of UX. Hence, general UX concepts can be derived based on AI-generated topics. Both pragmatic and hedonic dimensions are captured. Mostly, the items are coherent with each other and fit the construct. Especially, functional topics are well generated. However, some weaknesses must be stated. The items differ quite strongly and are accordingly not representative of the respective topic within some categories, e.g. **Identity**. Moreover, items (4) and (5) categorized in **Consistency and Integration** must be mentioned. The items are clearly of hedonic quality whereas the categorization and other items within the topic are considered pragmatic. Hence, there is a semantic relation between obviously functional and emotional items. For illustration, we have added a **(+)** for a suitable item fit and a **(-)** for an unsuitable item fit in the generated list (see Appendix A3). Additionally, some items may be contained in multiple topics. This can be traced back to the rather general formulation of the measurement items. If this was the case, we added **(+-)**.

*G. Prompt7: Searching for Items*

Up to this point, we showed how GenAI can be used to exploratively define a semantic structure on a large set of items. Another quite natural use case is to detect those items that represent a clearly defined UX concept. We demonstrate this in the example of the UX concept of learnability (or perspicuity). This concept describes that it is easy to get familiar with a product, i.e. easy to learn and understand how the product can be used [6]. We defined the following prompt *"Below there is a list of statements and questions related to the UX of a software system. Select all statements or questions from this list that describe how easy or difficult it is to learn and understand how to use the software system. List these statements or questions. Start with those statements and questions that describe this best.—inserted list of 408 items from UX questionnaires."*, i.e. an explanation of what we want plus the list of items used as a basis for the analysis.

The resulting list of items contained items that refer to ease of learning *("It was easy to learn to use this system")*, intuitive understanding *("The system was easy to use from the start")*, or aspects that support the user to handle the product *("Whenever I made a mistake using the system, I could recover easily and quickly")*. The top 15 of the resulting items fitted quite well to the request in the prompt (see Appendix A4). Thus, it is relatively simple to use ChatGPT to search for existing items that reflect certain UX concepts. Results indicate a good detection of relevant measurement items concerning the respective UX construct.

VI. CONCLUSION AND FUTURE WORK

This article presents a GenAI-based approach for providing a common ground in UX research. We applied the LLM ChatGPT-4 to analyze measurement items concerning semantic similarity from a pool of 408 items related to the most established UX questionnaires. Based on this, ChatGPT-4 generated generalized topics, subtopics, and the respective items. Lastly, ChatGPT detected representative items of existing UX concepts. As a result, six main topics and 15 subtopics were identified. In the following, theoretical and practical implications are drawn.

*A. Implications*

To conclude, we showed that LLMs can be used to (1) classify items from UX questionnaires concerning their semantic meaning, (2) improve and compare classifications, and (3) detect and assign items to classified topics. Of course, LLMs are inherently non-deterministic models. Thus, if the same sequence of prompts is used again, the resulting classifications will differ. This is in principle not a problem since there is no objectively "correct" classification. If the same task is done independently by several UX experts, the resulting classifications would of course differ too. However, the effort of such an automatic classification is extremely low, and thus the possibility to automatically create several such classifications allows an explorative search for semantic structures in large sets of items that can uncover interesting hidden dependencies that would be hard to detect with a manual analysis by UX experts.

Considering the results, ChatGPT generated a consolidated list of topics, subtopics, and items representing the concept UX comprehensively. Therefore, both functional and emotional aspects were contained. The AI-generated topics indicate a good alignment compared to existing UX concepts. In addition, ChatGPT detected and assigned suitable items to similar topics.

*B. Limitations and Future Research*

A severe limitation of the paper is that semantic differentials, a quite common item format in UX questionnaires, must be excluded from the analysis to guarantee at least a low level of comparability of the items. Further investigations in prompt engineering must show if it is possible to allow a combination of all common item formats in one analysis.

From a more practical point of view, the results can be used as a measurement framework for quantitative UX evaluation. In future research, a questionnaire for the holistic evaluation of the UX can be compiled from the AI-generated topics and the respective items. Moreover, items from the existing pool could be detected in relation to existing UX concepts and a comprehensive item list for each UX quality aspect can be set up. Such a list for each UX concept can help UX researchers by providing suitable measurement items quickly and easily. Both the questionnaire and the items could be further validated to compromise valid, reliable, and useful results.

This approach is a further step towards a common ground in UX research on the level of the measurement items.

## APPENDIX

**A1: Respective first three allocated items of AI-generated topics prompt1:**

**Usability and Ease of Use**
The system is easy to use.
I found the system unnecessarily complex.
I thought the system was easy to use.

**Design and Aesthetics**
The design is uninteresting.
The design appears uninspired.
The color composition is attractive.

**User Engagement and Experience**
I felt calm using the system.
I was so involved in this experience that I lost track of time.
I lost myself in this experience.

**Trust and Reliability**
I feel comfortable purchasing from the website.
I feel confident conducting business on the website.
It is a site that feels secure.

**Information Access and Clarity**
I am able to get the information I need easily.
provides quick and easy access to finding information.
provides relevant information.

**Issues and Errors**
The system is too inflexible.
The interaction with the system is irritating.
The interaction with the system is frustrating.

**A2: Respective first three allocated items of AI-generated topics of prompt2:**

**Ease of Use**
The system is easy to use.
I thought the system was easy to use.
This system is easy to use.

**Complexity and Usability Issues**
I found the system unnecessarily complex.
I think that I would need the support of a technical person to be able to use this system.
I found this Application X confusing to use.

**Design and Appearance**
The design is uninteresting.
The design appears uninspired.
The color composition is attractive.

**Engagement and Immersion**
I felt calm using the system.
I was so involved in this experience that I lost track of time.
I lost myself in this experience.

**Performance and Responsiveness**
The software responds too slowly.
This software responds too slowly to inputs.
The interaction with the system is fast.

**Reliability and Trust**
I feel comfortable purchasing from the website.
I feel confident conducting business on the website.
It is a site that feels secure.

**Information Quality and Access**
I am able to get the information I need easily.
provides quick and easy access to finding information.
provides relevant information.

**Errors and Bugs**
The system is too inflexible.
The interaction with the system is irritating.
The software has at some time stopped unexpectedly.

**Learning and Memorability**
I think that I would need the support of a technical person to be able to use this system.
I easily remember how to use it.
It is easy to learn to use it.

**Effectiveness and Efficiency**
It helps me be more effective.
It meets my needs.
It does everything I would expect it to do.

**A3: Generalized AI-generated UX topics with their respective top-5 items:**

**Usability and Intuitiveness**

Ease of Use:
1) The system is easy to use. (+)
2) I think that I would need the support of a technical person to be able to use this system.(+)
3) I would imagine that most people would learn to use this system very quickly.(+)
4) I learned to use it quickly.(+)
5) I can use it without written instructions.(+)

Efficiency and Speed:
1) The interaction with the system is fast.(+)
2) The system responds too slowly.(+)
3) This software responds too slowly to inputs.(+)
4) The speed of this software is fast enough.(+)
5) has fast navigation to pages.(+)

Adaptability:
1) The system is too inflexible.(+)
2) This software seems to disrupt the way I normally like to arrange my work.(+)
3) It is flexible.(+)
4) It requires the fewest steps possible to accomplish what I want to do with it.(+- Efficiency)
5) It is relatively easy to move from one part of a task to another.(+- Efficiency)

**Content Quality and Clarity**

Relevance and Utility:
1) Provides relevant information.(+)
2) It meets my needs.(+)
3) It is useful.(+)
4) Provides information content that is easy to read.(+)
5) It does everything I would expect it to do.(+)

Consistency and Integration:
1) I thought there was too much inconsistency in this system.(+)
2) I found the various functions in this system were well integrated.(+)
3) I don't notice any inconsistencies as I use it.(+)
4) Everything goes together on this site.(+-)
5) The site appears patchy.(+-)

Clarity and Understandability:
1) The way that system information is presented is clear and understandable.(+)
2) provides information content that is easy to understand.(+)
3) I think the image is difficult to understand.(+)
4) The layout is easy to grasp.(+)
5) I do not find this image useful.(-)

**Engagement and Experience**

Engagement Level:
1) I was so involved in this experience that I lost track of time.(+)
2) I lost myself in this experience.(+)
3) I was really drawn into this experience.(+)
4) I felt involved in this experience.(+)
5) I was absorbed in this experience.(+)

Stimulation:
1) This experience was fun.(+)
2) I continued to use Application X out of curiosity.(+)
3) Working with this software is mentally stimulating.(+)
4) I felt involved in this experience.(+)
5) During this experience I let myself go.(+- Engagement Level)

Aesthetics and Design:
1) This Application X was aesthetically appealing.(+)
2) The screen layout of Application X was visually pleasing.(+)
3) The design is uninteresting.(+)
4) The layout appears professionally designed.(+)
5) The design appears uninspired.(+)

**Trust and Reliability**

Trust and Security:
1) I feel comfortable purchasing from the website.(+)
2) I feel confident conducting business on the website.(+)
3) is a site that feels secure.(+)
4) makes it easy to contact the organization.(+)
5) The website is easy to use.(-)

Dependability:
1) This software hasn't always done what I was expecting.(+)
2) The software has helped me overcome any problems I have had in using it.(+)
3) I can recover from mistakes quickly and easily.(+)
4) I can use it successfully every time.(+)
5) Error messages are not adequate.(+)

**Novelty and Identity**

Novelty:
1) The layout is inventive.(+)
2) The layout appears dynamic.(-)
3) The layout appears too dense.(-)
4) The layout is pleasantly varied.(-)
5) The design of the site lacks a concept.(-)

Identity:
1) Conveys a sense of community.(+)
2) The offer has a clearly recognizable structure.(-)
3) Keeps the user's attention.(-)
4) The layout is not up-to-date.(-)
5) The design of the site lacks a concept.(-)

**Value and Loyalty**

Perceived Value:
1) I consider my experience a success.(+)
2) My experience was rewarding.(+)
3) The layout appears professionally designed.(+)
4) The color composition is attractive.(+)
5) It is wonderful.(+)

Loyalty:
1) I would recommend Application X to my family and friends.(+)
2) I would recommend this software to my colleagues.(+)
3) I will likely return to the website in the future.(+)
4) I think that I would like to use this system frequently.(+)
5) I would not want to use this image.(+)

**A4: Top 15 items filtered for Perspicuity/Learnability**
1) It was easy to learn to use this system
2) I could effectively complete the tasks and scenarios using this system
3) I was able to complete the tasks and scenarios quickly using this system
4) I felt comfortable using this system
5) The system gave error messages that clearly told me how to fix problems
6) Whenever I made a mistake using the system, I could recover easily and quickly
7) The information provided with this system (online help, documentation) was clear
8) It was easy to find the information I needed
9) The information provided for the system was easy to understand
10) The information was effective in helping me complete the tasks and scenarios
11) The system was easy to use from the start
12) How the system is used was clear to me straight away
13) I could interact with the system in a way that seemed familiar to me
14) It was always clear to me what I had to do to use the system
15) The process of using the system went smoothly